\documentclass{iaus}
\usepackage{graphicx}
\usepackage{caption}

\title[JD 11.~~Massive eclipsing binaries in Westerlund 1] 
{Fundamental Parameters of 4 Massive Eclipsing Binaries in Westerlund 1}

\author[Koumpia \& Bonanos]   
{E. Koumpia
 \and A.Z. Bonanos
\footnote[1]{EK \& AZB acknowledge support from the IAU and the European Commission for an FP7 Marie Curie International Reintegration Grant.}}

\affiliation{National Observatory of Athens, Institute of Astronomy \& Astrophysics, \\ I. Metaxa \& Vas. Pavlou St., Palaia Penteli GR-15236 Athens, Greece \\[\affilskip]
{\tt koumpia@astro.noa.gr}, {\tt bonanos@astro.noa.gr} \\[\affilskip]}

\pubyear{2010}
\volume{272}  
\pagerange{1--3}
\jname{Active OB stars: structure, evolution, mass loss \\ and critical limits}
\editors{C. Neiner, G. Wade, G. Meynet \& G. Peters}
\begin{document}

\maketitle

\begin{abstract}
Westerlund 1 is one of the most massive young clusters known in the Local Group, with an age of 3-5 Myr. It contains an assortment of rare evolved massive stars, such as blue, yellow and red supergiants, Wolf-Rayet stars, a luminous blue variable, and a magnetar, as well as 4 massive eclipsing binary systems (Wddeb, Wd13, Wd36, WR77o, see
\cite[Bonanos 2007]{Bonanos07}). 
The eclipsing binaries present a rare opportunity to constrain evolutionary models of massive stars, the distance to the cluster and furthermore, to determine a dynamical lower limit for the mass of a magnetar progenitor. Wddeb, being a detached system, is of great interest as it allows determination of the masses of 2 of the most massive unevolved stars in the cluster. We have analyzed spectra of all 4 eclipsing binaries, taken in 2007-2008 with the 6.5 meter Magellan telescope at Las Campanas Observatory, Chile, and present fundamental parameters (masses, radii) for their component stars. 
\end{abstract}

\textbf{1. Introduction}

Westerlund 1 (Wd1) is one of the most massive compact young star clusters known in the Local Group. It was discovered by
\cite[Westerlund (1961)]{Westerlund61},
but remained largely unstudied for many years due to high interstellar extinction in its direction. The cluster contains a large number of rare, evolved high-mass stars including 6 yellow hypergiants, 4 red supergiants, 24 Wolf-Rayet stars, a luminous blue variable and many OB supergiants. In addition, X-ray observations have revealed the presence of the magnetar CXO J164710.2-455216, a slow X-ray pulsar that must have formed from a high-mass progenitor star
(\cite[Muno et al. 2006]{Muno_etal06}).
Wd1 is believed to have formed in a single burst of star formation, implying the constituent stars are of the same age and composition.

The study of eclipsing binaries in this cluster is important for several reasons: (1) the determination of fundamental parameters of the component stars (mass, radii, etc.), in order to test the evolutionary models of massive stars, (2) EBs provide an independent measurement of the distance, based on the expected absolute magnitude, and (3) the determination of a dynamical lower limit for the mass of the magnetar progenitor.

\textbf{2. The individual binaries}

We have analyzed spectra of 4 eclipsing binaries taken in 2007-2008 with the 6.5 meter Magellan telescope at Las Campanas Observatory, Chile. The spectra were reduced and extracted using IRAF\footnote[2]{IRAF is distributed by the NOAO, which are operated by the Association of Universities for Research in Astronomy, Inc., under cooperative agreement with the NSF.}. We used two methods to determine the radial velocities: Gaussian fitting of the line cores with the IRAF $noao.rv.rvidlines$ package and $\chi^2$ minimization, which finds the least $\chi^2$ from the observed spectrum and fixed synthetic TLUSTY models 
(\cite[Lanz \& Hubeny 2003]{LanzHubeny03}).
In both cases, we chose to use the narrow Helium lines ($\lambda\lambda6678,7065$), as they are less sensitive to systematics, rather than the broader hydrogen lines. We adopted the velocities resulting from the second method as they seemed to be more robust. 

Our preliminary results for the parameters of the members of the 4 systems (period, masses, radii, surface gravities of the components, eccentricity and inclination), as modelled using PHOEBE software
(\cite[Prsa \& Zwitter 2005]{PrsaZwitter05}),
are presented in Table \ref{tab1}.

\begin{center}
{\footnotesize
	\begin{tabular}{|l|c|c|c|c|c|c|c|c|c|c|}
		\textbf{Binary} & \textbf{P(days)} & \textbf{M$_1$(\mbox{M$_{\odot}$})} & \textbf{M$_2$(\mbox{M$_{\odot}$})} & \textbf{R$_1$(\mbox{R$_{\odot}$})} & \textbf{R$_2$(\mbox{R$_{\odot}$})} & \textbf{logg$_1$} & \textbf{logg$_2$} & \textbf{eccentr} & \textbf{Incl($^o$)}\\ \hline
		\textbf{Wddeb} & 4.447 & 9.62 & 13.84 & 5.47 & 6.22 & 3.94 & 3.99 & 0.177 & 84.46 \\ \hline
		\textbf{Wd36} & 3.182 & 10.75 & 12.51 & 9.45 & 10.15 & 3.52 & 3.52 & 0(fixed) & 72.72 \\ \hline
		\textbf{WR77o} & 3.520 & 49.67 & 19.87 & 18.06 & 11.81 & 3.62 & 3.59 & 0(fixed) & 58.53 \\ \hline
		\textbf{Wd13} & 9.267 & 29.91 & 23.86 & 26.88 & 24.08 & 3.05 & 3.05 & 0(fixed) & 55.08 \\ 
	\end{tabular}
}
	\firstsection
	\captionof{table}{Fundamental Parameters of Eclipsing Binaries in Wd1}
	\label{tab1}
\end{center}

\textbf{\underline{Wd13}}: is a semi-detached, double-lined spectroscopic binary (B0.5Ia$^{+}$ and OB types,
\cite[Negueruela et al. 2010]{Negueruela_etal10}) in a circular orbit. We can see both absorption and emission lines in its spectra.\\
\textbf{\underline{Wd36}}: is an overcontact, double-lined spectroscopic binary system (OB-type) in a circular orbit (see Fig.\,\ref{fig1}).\\
\textbf{\underline{WR77o}}: is probably a double contact binary system in an almost circular orbit. It is a single lined spectroscopic binary and thus it is difficult to determine the parameters of its component stars. The spectroscopic visible star is a Wolf-Rayet star of spectral type WN6-7 
(\cite[Negueruela \& Clark 2005]{NegueruelaClark05}), 
given the large line widths of 2000 km~s$^{-1}$.\\
\textbf{\underline{Wddeb}}: is a detached, double-lined spectroscopic binary system (OB-type) with an eccentricity of almost 0.2.

\begin{center}
	\includegraphics[scale=0.23]{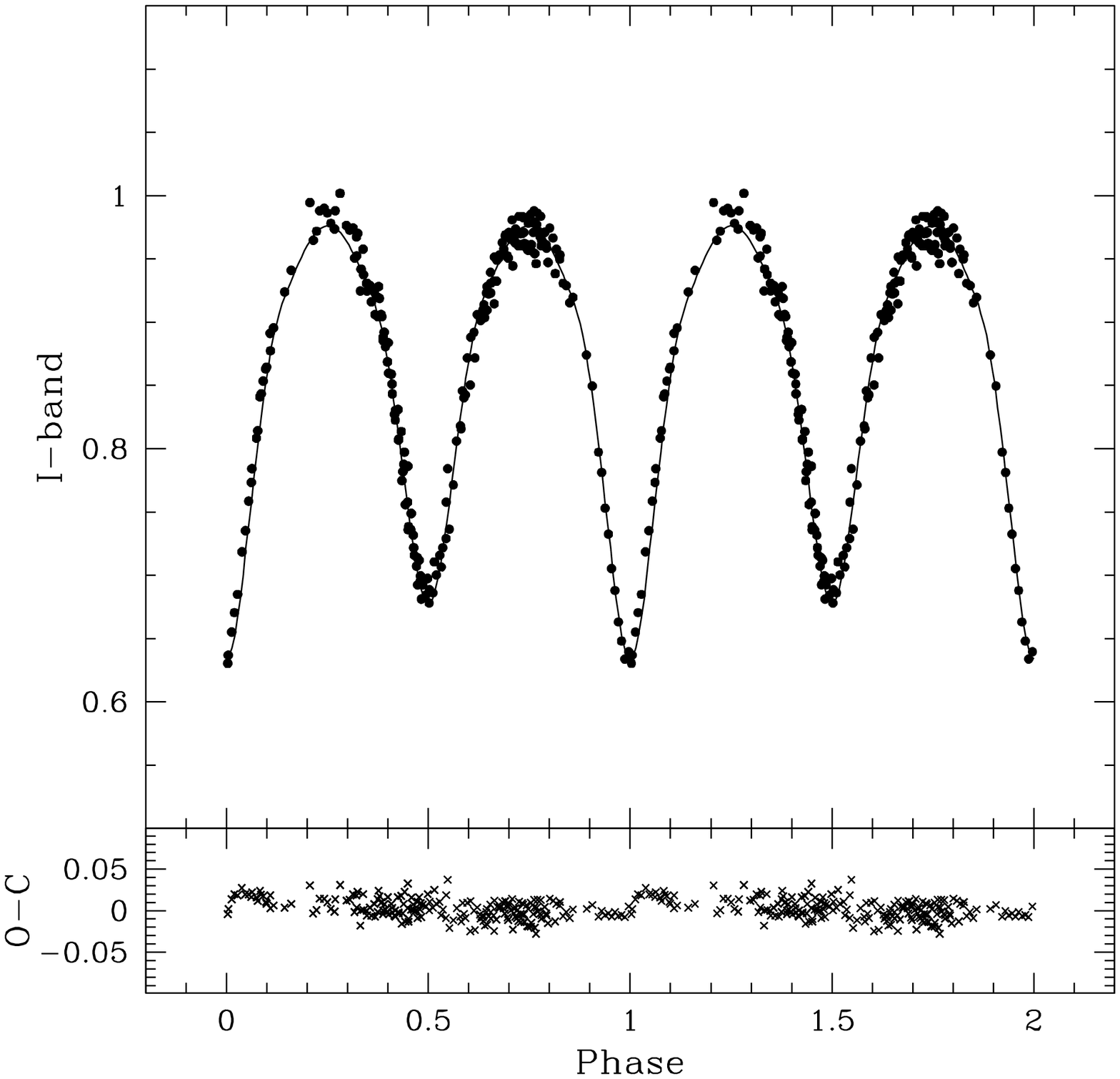}\includegraphics[scale=0.23]{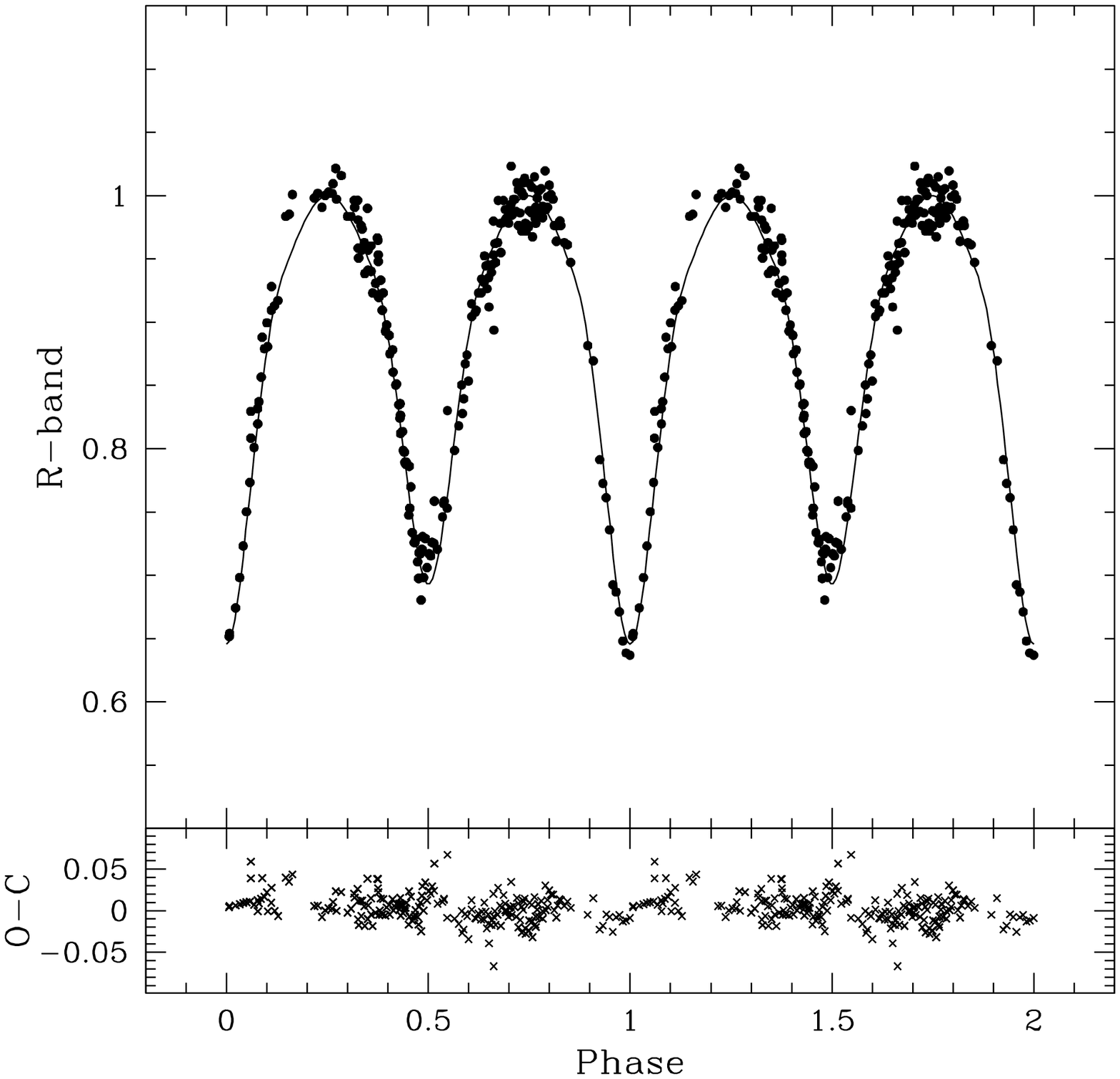}\includegraphics[scale=0.23]{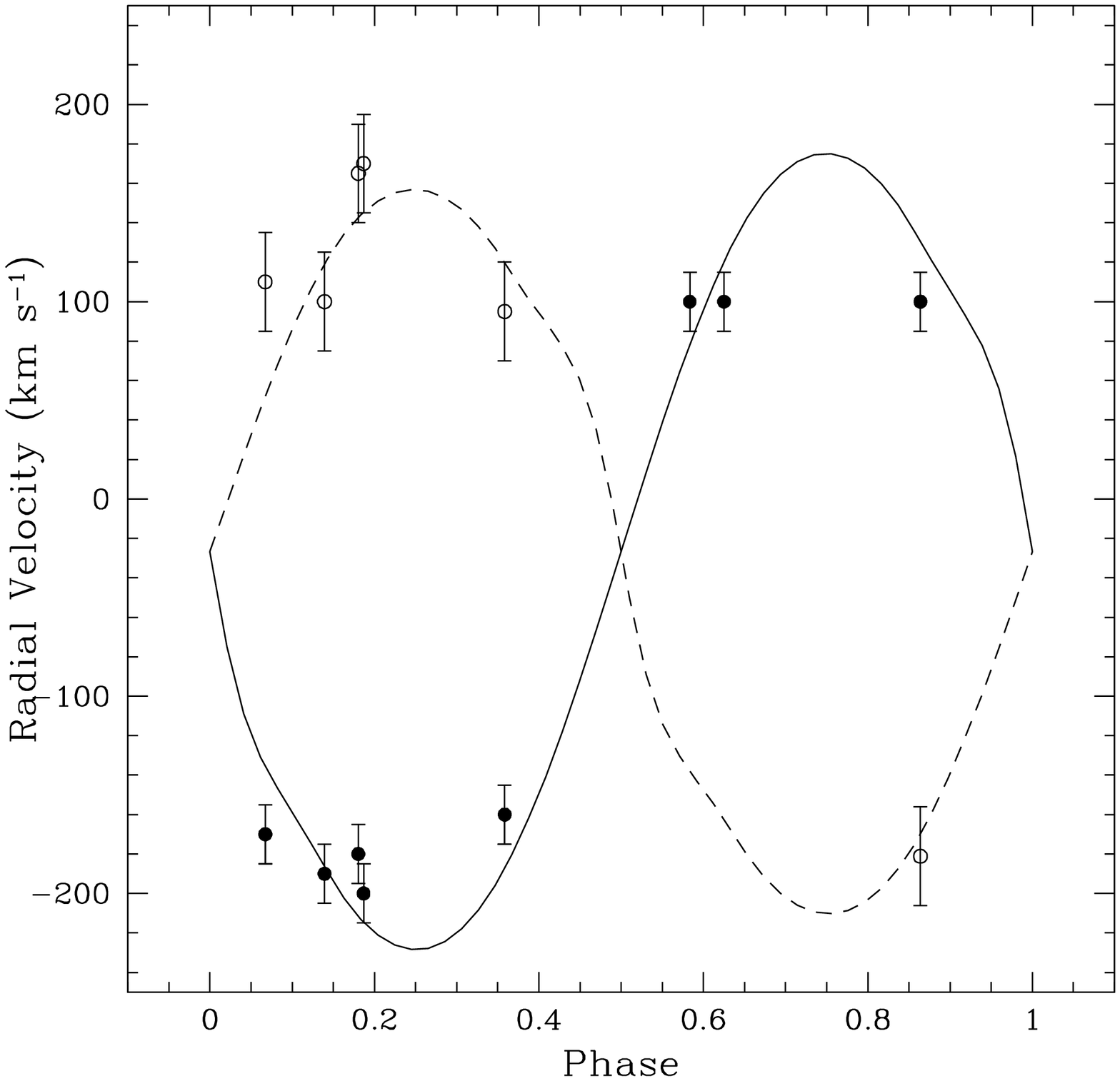}
	\firstsection
 \captionof{figure}{Phased $I,R$ light curves and radial velocity curve of Wd36, respectively.}
   	\label{fig1}
\end{center}

\firstsection

\end{document}